\documentclass[aps,showkeys,showpacs,twocolumn,showpacs,preprintnumbers,amsmath,amssymb,superscriptaddress,floatfix,nofootinbib]{revtex4}
\usepackage{graphicx,color,dcolumn,booktabs,bm}
\usepackage{longtable,lscape}
\usepackage{txfonts}
\usepackage{overpic}
\usepackage{epsfig}
\usepackage{amssymb}
\usepackage{rotating}
\usepackage{epstopdf}
\usepackage{indentfirst}
\usepackage{feynmf}   %{feynmp}
\usepackage{slashed}  %for Feynman symbols
\usepackage{cases}
\usepackage{color}
\usepackage{multirow}
\usepackage{graphicx,color,dcolumn,booktabs,bm}
\usepackage{cases}
\usepackage{array}

\graphicspath{{Figures/}} %
\usepackage[colorlinks, citecolor=blue,anchorcolor=red,menucolor=red, linkcolor=red,filecolor=red,runcolor=red,urlcolor=blue,frenchlinks=red]{hyperref}
\begin{document}

\title{Canonical interpretation of the newly observed $\Lambda_b(6146)^0$ and $\Lambda_b(6152)^0$ via strong decay behaviors}

\author{Wei Liang}
\affiliation{  Department
of Physics, Hunan Normal University,  Changsha 410081, China }

\affiliation{ Synergetic Innovation
Center for Quantum Effects and Applications (SICQEA), Changsha 410081,China}

\affiliation{  Key Laboratory of
Low-Dimensional Quantum Structures and Quantum Control of Ministry
of Education, Changsha 410081, China}

\author{Qi-Fang L\"{u}} \email[]{Corresponding author: lvqifang@hunnu.edu.cn}
\affiliation{  Department
of Physics, Hunan Normal University,  Changsha 410081, China }

\affiliation{ Synergetic Innovation
Center for Quantum Effects and Applications (SICQEA), Changsha 410081,China}

\affiliation{  Key Laboratory of
Low-Dimensional Quantum Structures and Quantum Control of Ministry
of Education, Changsha 410081, China}

\author{Xian-Hui Zhong} \email[]{Corresponding author: zhongxh@hunnu.edu.cn}
\affiliation{  Department
of Physics, Hunan Normal University,  Changsha 410081, China }

\affiliation{ Synergetic Innovation
Center for Quantum Effects and Applications (SICQEA), Changsha 410081,China}

\affiliation{  Key Laboratory of
Low-Dimensional Quantum Structures and Quantum Control of Ministry
of Education, Changsha 410081, China}
\begin{abstract}

Stimulated by the newly observed $\Lambda_b(6146)^0$ and $\Lambda_b(6152)^0$ resonances, we investigate the strong decays of the low-lying singly bottom baryons within the quark pair creation model. Considering their masses and decay modes, we tentatively assign these states as the $\lambda-$ mode $\Lambda_b(2S)$, $\Lambda_b(1D)$, $\Sigma_b(2S)$, and $\Sigma_b(1P)$ states, and then estimate their strong decay widths. Compared with the experimental total widths and decay modes measured by the LHCb Collaboration, the $\Lambda_b(6146)^0$ and $\Lambda_b(6152)^0$ can be reasonably clarified into the $J^P=5/2^+$ and $J^P=3/2^+$ $\Lambda_b(1D)$ sates respectively, while other canonical assignments are disfavored. Other low-lying bottom states are also presented, where various narrow states may have good potential to be observed in future LHCb experiments.
\end{abstract}


\maketitle

\section{Introduction}{\label{introduction}}

Very recently, the LHCb Collaboration reported the discovery of two new singly bottom baryons $\Lambda_b(6146)^0$ and $\Lambda_b(6152)^0$ in the $\Lambda_b^0 \pi^+ \pi^-$ invariant mass distribution~\cite{lhcb}. Their measured masses and widths are listed as follows,
\begin{eqnarray}
m[\Lambda_b(6146)^0] = 6146.17\pm0.33\pm0.22\pm0.16~\rm{MeV},
\end{eqnarray}
\begin{eqnarray}
\Gamma[\Lambda_b(6146)^0] = 2.9\pm1.3\pm0.3~\rm{MeV},
\end{eqnarray}
\begin{eqnarray}
m[\Lambda_b(6152)^0] = 6152.51\pm0.26\pm0.22\pm0.16~\rm{MeV},
\end{eqnarray}
\begin{eqnarray}
\Gamma[\Lambda_b(6152)^0] = 2.1\pm0.8\pm0.3~\rm{MeV},
\end{eqnarray}
where the mass gap between these two states are only about 6 MeV. Compared with the masses predicted by the constituent quark model~\cite{Capstick:1986bm,Ebert:2007nw,Roberts:2007ni,Ebert:2011kk,Chen:2014nyo}, the LHCb Collaboration suggested that these two states can be assigned as the $1D$ doublet of the $\Lambda_b^0$ states~\cite{lhcb}.

A singly bottom baryon without strange quark is composed of one bottom quark and two light quarks. These bottom baryons can be categorized into the $\Lambda_b$ and $\Sigma_b$ families, which belong to the antisymmetric flavor configuration $\bar 3_F$ and symmetric flavor configuration $6_F$, respectively. From the Review of Particle Physics~\cite{Tanabashi:2018oca}, there exist six $\Lambda_b$ and $\Sigma_b$ baryons, $\Lambda_b(5620)$, $\Lambda_b(5912)$, $\Lambda_b(5920)$, $\Sigma_b(5811)$, $\Sigma_b^*(5830)$, and $\Sigma_b(6097)$. The $\Lambda_b(5620)$, $\Sigma_b(5811)$, and $\Sigma_b^*(5830)$ are the $S-$wave ground states undoubtedly. The $\Lambda_b(5912)$ and $\Lambda_b(5920)$ can be well established as the two $\lambda-$mode $P-$wave $\Lambda_b$ states by various theoretical studies~\cite{Capstick:1986bm,Ebert:2007nw,Roberts:2007ni,Ebert:2011kk,Chen:2014nyo,Garcilazo:2007eh,Karliner:2015ema,Mao:2015gya,Thakkar:2016dna,Wang:2017kfr}. Under the canonical interpretation, the mass and strong decay behaviors indicate that $\Sigma_b(6097)$ can be categorized into the $\lambda-$mode $P-$wave $\Sigma_b$ states in the literature~\cite{Chen:2018vuc,Wang:2018fjm,Yang:2018lzg,Aliev:2018vye,Cui:2019dzj,Jia:2019bkr}. Due to lack of more experimental information, the spectra of these low-lying bottom baryons have not been established so far~\cite{Chen:2016spr,Crede:2013sze,Cheng:2015iom,Richard:1992uk,Klempt:2009pi}.

Although the LHCb Collaboration suggested that these two states seem to be the $\Lambda_b(1D)$ doublet, other interpretations are possible. In fact, because these two states are observed in the $\Lambda_b^0 \pi^+ \pi^-$ final states, they probably correspond to the neutral excited $\Sigma_b$ states~\cite{lhcb}. Hence, the newly observed two states have the potential to be the $\Sigma_b(6146)^0$ and $\Sigma_b(6152)^0$ resonances, and we also need to examine this $I=1$ case. Moreover, the constituent quark model may have significant uncertainties on the mass spectrum, and just perform the mass pattern of these canonical baryon states. Considering the mass uncertainties predicted by the quark model, we find that these two resonances lie in the mass region of the predicted $\lambda-$ mode $\Lambda_b(2S)$, $\Lambda_b(1D)$, $\Sigma_b(2S)$, and $\Sigma_b(1P)$ states. To clarify their nature, the strong decays of the these assignments should be examined carefully.

The studies on these low-lying bottom baryons can help us to understand the nature of these two newly observed resonances and establish the bottom baryon spectrum. Before this discovery by the LHCb Collaboration, the strong decays of the low-lying bottom baryons have been investigated by several works. However,  due to the lack of the experimental information, the choices of parameters may be ambiguous and the predictions of these different works do not agree with each other. Hence, it is essential to investigate the strong decays of these low-lying bottom baryons within the unified model and coincident parameters. In this issue, we calculate the strong decay behaviors of these states within the $^3P_0$ model systematically. Our results shows that the $\Lambda_b(6146)^0$ and $\Lambda_b(6152)^0$ can be assigned as the $\Lambda_b(1D)$ doublet reasonably, while other canonical assignments are disfavored. The predictions of other low-lying bottom states may provide helpful information for future experimental searches.

This paper is organized as follows. The $^3P_0$ model and notations are introduced in Sec.~\ref{model}. The strong decays of the low-lying bottom baryons are estimated in Sec.~\ref{decay}. A short summary is presented in the last section.

\section{$^3P_0$ Model}{\label{model}}

In this calculation, the $^3P_0$ quark pair creation model is adopted to investigate the
Okubo-Zweig-Iizuka-allowed two-body strong decays of the excited $\Lambda_b$ and $\Sigma_b$ states. The $^3P_0$ model has been widely employed to estimate the strong decay behaviors of conventional hadrons and achieved with considerable successes~\cite{micu,3p0model1,3p0model2,3p0model4,3p0model5,3p0model6,Chen:2007xf,Zhao:2016qmh,Ye:2017dra,Chen:2017gnu,Lu:2014zua,
Lu:2016bbk,Ferretti:2014xqa,Godfrey:2015dva,Segovia:2012cd,Mu:2014iaa,Lu:2018utx,Guo:2019ytq,Chen:2016iyi}. In this model, the initial hadrons can decay into two final states via a quark-antiquark pair with the vacuum quantum number $J^{PC}=0^{++}$~\cite{micu}. For the baryons, the transition operator $T$ of the decay $A\rightarrow BC$ can be written as~\cite{Chen:2007xf,Chen:2017gnu}
\begin{eqnarray}
T&=&-3\gamma\sum_m\langle 1m1-m|00\rangle\int
d^3\boldsymbol{p}_4d^3\boldsymbol{p}_5\delta^3(\boldsymbol{p}_4+\boldsymbol{p}_5)\nonumber\\
&&\times {\cal{Y}}^m_1\left(\frac{\boldsymbol{p}_4-\boldsymbol{p}_5}{2}\right
)\chi^{45}_{1,-m}\phi^{45}_0\omega^{45}_0b^\dagger_{4i}(\boldsymbol{p}_4)d^\dagger_{4j}(\boldsymbol{p}_5),
\end{eqnarray}
where $\gamma$ is a dimensionless constant reflecting the $q_4\bar{q}_5$ pair-production strength, and $\boldsymbol{p}_4$ and
$\boldsymbol{p}_5$ are the momenta of the created quark $q_4$ and
antiquark $\bar{q}_5$, respectively. The solid harmonic polynomial
${\cal{Y}}^m_1(\boldsymbol{p})\equiv|p|Y^m_1(\theta_p, \phi_p)$ stands
the $P-$wave momentum-space distribution of the created quark pair. $\phi^{45}_{0}=(u\bar u + d\bar d +s\bar s)/\sqrt{3}$,
$\omega^{45}=\delta_{ij}$, and $\chi_{{1,-m}}^{45}$ are the flavor singlet, color singlet,
and spin triplet wave functions of the  $q_4\bar{q}_5$, respectively. The $b^\dagger_{4i}(\boldsymbol{p}_4)d^\dagger_{4j}(\boldsymbol{p}_5)$ corresponds to the creation operators of the created quark and antiquark, where the $i$ and $j$ are the color indices.

For the relevant hadrons, the definition of the mock states should be adopted. Take the initial baryon $A$ as an example, the total wave function can be expressed as~\cite{Hayne:1981zy}
\begin{eqnarray}
&&|A(n^{2S_A+1}_AL_{A}\,\mbox{}_{J_A M_{J_A}})(\boldsymbol{P}_A)\rangle
\equiv \nonumber\\
&& \sqrt{2E_A}\sum_{M_{L_A},M_{S_A}}\langle L_A M_{L_A} S_A
M_{S_A}|J_A
M_{J_A}\rangle \int d^3\boldsymbol{p}_1d^3\boldsymbol{p}_2d^3\boldsymbol{p}_3\nonumber\\
&&\times \delta^3(\boldsymbol{p}_1+\boldsymbol{p}_2+\boldsymbol{p}_3-\boldsymbol{P}_A)\psi_{n_AL_AM_{L_A}}(\boldsymbol{p}_1,\boldsymbol{p}_2,\boldsymbol{p}_3)\chi^{123}_{S_AM_{S_A}}
\phi^{123}_A\omega^{123}_A\nonumber\\
&&\times  \left|q_1(\boldsymbol{p}_1)q_2(\boldsymbol{p}_2)q_3(\boldsymbol{p}_3)\right\rangle,
\end{eqnarray}
which satisfies the normalization condition
\begin{eqnarray}
\langle A(\boldsymbol{P}_A)|A(\boldsymbol{P}^\prime_A)\rangle=2E_A\delta^3(\boldsymbol{P}_A-\boldsymbol{P}^\prime_A).
\end{eqnarray}
Here, the $\boldsymbol{P}_A$, $\boldsymbol{p}_1$, $\boldsymbol{p}_2$, and $\boldsymbol{p}_3$ are the momenta of the $A$, $q_1$, $q_2$, and $q_3$, respectively. $E_A$ is the total energy of the baryon $A$. $\chi^{123}_{S_AM_{S_A}}$, $\phi^{123}_A$, $\omega^{123}_A$, $\psi_{n_AL_AM_{L_A}}(\boldsymbol{p}_1,\boldsymbol{p}_2,\boldsymbol{p}_3)$ are the spin, flavor, color, and
space wave functions of the baryon $A$, respectively. The definitions of the mock states $B$ and $C$ are similar to that of initial state $A$, and can be found in Ref.~\cite{Chen:2007xf}.

\begin{figure*}[!htbp]
\includegraphics[scale=1.3]{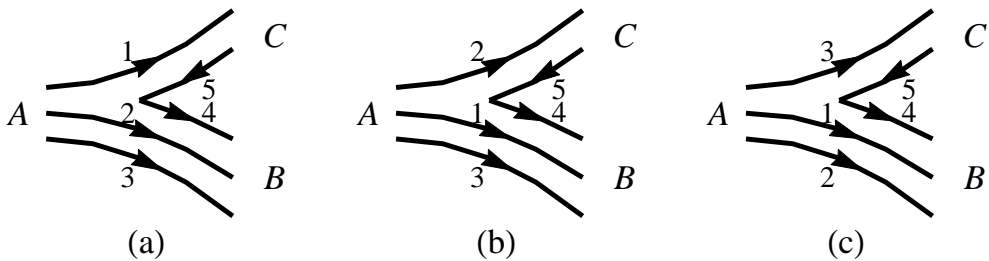}
\vspace{0.0cm} \caption{The baryon decay process $A\to B+C$ in the $^3P_0$ model.}
\label{qpc}
\end{figure*}

For the decay of the singly bottom baryon $A$, there are three possible rearrangements,
\begin{eqnarray}
A(q_1,q_2,b_3)+P(q_4,\bar q_5)\to B(q_2,q_4,b_3)+C(q_1,\bar q_5),\\
A(q_1,q_2,b_3)+P(q_4,\bar q_5)\to B(q_1,q_4,b_3)+C(q_2,\bar q_5),\\
A(q_1,q_2,b_3)+P(q_4,\bar q_5)\to B(q_1,q_2,q_4)+C(b_3,\bar q_5),
\end{eqnarray}
where the $q_i$ and $b_3$ denote the light quark and bottom quark, respectively. These three ways of recouplings are also shown in Figure~\ref{qpc}. It can be seen that the first and second cases stand for the heavy baryon plus the light meson decay modes, while the last one corresponds to the light baryon plus the heavy-light meson final states. In our present calculations, the light baryon plus the heavy-light meson decay mode do not exist due to the phase space constraint.

The $S$ matrix is defined as
\begin{eqnarray}
\langle
f|S|i\rangle=I-i2\pi\delta(E_f-E_i){\cal{M}}^{M_{J_A}M_{J_B}M_{J_C}},
\end{eqnarray}
where the ${\cal{M}}^{M_{J_A}M_{J_B}M_{J_C}}$ corresponds to the helicity amplitude of the decay process $A\to B+C$. Given the $A(q_1,q_2,b_3)+P(q_4,\bar q_5)\to B(q_1,q_4,b_3)+C(q_2,\bar q_5)$ shown in Fig. 1(b), the helicity amplitude ${\cal{M}}^{M_{J_A}M_{J_B}M_{J_C}}$ can be expressed as~\cite{Chen:2007xf,Ye:2017dra},
\begin{eqnarray}
&&\delta^3(\boldsymbol{p}_B+\boldsymbol{p}_C-\boldsymbol{p}_A){\cal{M}}^{M_{J_A}M_{J_B}M_{J_C}} = \nonumber\\
&&- \gamma \sqrt{8E_AE_BE_C} \sum_{M_{\rho_A}} \sum_{M_{L_A}} \sum_{M_{\rho_B}} \sum_{M_{L_B}} \sum_{M_{S_1}, M_{S_3}, M_{S_4}, m}\nonumber\\
&& \times  \langle j_A M_{j_A}S_3M_{S_3}|J_AM_{J_A}\rangle \langle L_{\rho_A} M_{L_{\rho_A}}L_{\lambda_A}M_{L_{\lambda_A}}|L_AM_{L_A} \rangle \nonumber\\ && \times \langle L_A M_{L_A}S_{12}M_{S_{12}}|j_AM_{j_A}\rangle \langle S_1M_{S_1}S_2M_{S_2}|S_{12}M_{S_{12}} \rangle \nonumber\\
&& \times \langle j_B M_{j_B}S_3M_{S_3}|J_BM_{J_B}\rangle \langle L_{\rho_B} M_{L_{\rho_B}}L_{\lambda_B}M_{L_{\lambda_B}}|L_BM_{L_B}\rangle \nonumber\\ && \times \langle L_B M_{L_B}S_{14}M_{S_{14}}|j_BM_{j_B}\rangle \langle S_1M_{S_1}S_4M_{S_4}|S_{14}M_{S_{14}}\rangle\nonumber\\
&& \times \langle 1m 1-m|00\rangle \langle S_4M_{S_4}S_5M_{S_5}|1-m \rangle \nonumber\\
&& \times \langle L_C M_{L_C}S_CM_{S_C}|J_CM_{J_C}\rangle \langle S_2M_{S_2}S_5M_{S_5}|S_CM_{S_C}\rangle \nonumber\\
&& \times \langle \phi_B^{143} \phi_C^{25}|\phi_A^{123}\phi_0^{45}\rangle I^{M_{L_A}m}_{M_{L_B}M_{L_C}}(\boldsymbol{p}),
\end{eqnarray}
where $\langle \phi_B^{143} \phi_C^{25}|\phi_A^{123}\phi_0^{45}\rangle$ is the overlap of the flavor wavefunctions, and the $I^{M_{L_A}m}_{M_{L_B}M_{L_C}}(\boldsymbol{p})$ is the spatial overlap of the initial and final states. Here, the $I^{M_{L_A}m}_{M_{L_B}M_{L_C}}(\boldsymbol{p})$ can be written as
\begin{eqnarray}
I^{M_{L_A}m}_{M_{L_B}M_{L_C}}(\boldsymbol{p}) & = & \int d^3\boldsymbol{p}_1d^3\boldsymbol{p}_2d^3\boldsymbol{p}_3d^3\boldsymbol{p}_4d^3\boldsymbol{p}_5  \nonumber\\ && \times \delta^3(\boldsymbol{p}_1+\boldsymbol{p}_2+\boldsymbol{p}_3-\boldsymbol{P}_A)\delta^3(\boldsymbol{p}_4+\boldsymbol{p}_5)\nonumber\\ && \times \delta^3(\boldsymbol{p}_1+\boldsymbol{p}_4+\boldsymbol{p}_3-\boldsymbol{P}_B)\delta^3(\boldsymbol{p}_2+\boldsymbol{p}_5-\boldsymbol{P}_C) \nonumber\\
&& \times \psi^*_B(\boldsymbol{p}_1,\boldsymbol{p}_4,\boldsymbol{p}_3) \psi^*_C(\boldsymbol{p}_2,\boldsymbol{p}_5)\nonumber\\
&& \times\psi_A(\boldsymbol{p}_1,\boldsymbol{p}_2,\boldsymbol{p}_3){\cal{Y}}^m_1\left(\frac{\boldsymbol{p}_4-\boldsymbol{p}_5}{2}\right
).
\end{eqnarray}
It should be noted that the $I^{M_{L_A}m}_{M_{L_B}M_{L_C}}(\boldsymbol{p})$ does not only depend on the magnetic quantum numbers, but also relies on the radial and orbital quantum numbers of the relevant hadrons that are usually omitted for simplicity in the literature.

In this work, the simplest vertex with
a spatially constant pair production strength $\gamma$, the relativistic phase space, and the simple harmonic oscillator wave functions are used as in Ref.~\cite{micu}. Then, the decay width
$\Gamma(A\rightarrow BC)$ can be calculated straightforward
\begin{eqnarray}
\Gamma= \pi^2\frac{p}{M^2_A}\frac{s}{2J_A+1}\sum_{M_{J_A},M_{J_B},M_{J_C}}|{\cal{M}}^{M_{J_A}M_{J_B}M_{J_C}}|^2,
\end{eqnarray}
where $p=|\boldsymbol{p}|=\frac{\sqrt{[M^2_A-(M_B+M_C)^2][M^2_A-(M_B-M_C)^2]}}{2M_A}$ is the momentum of the final hadrons. The statistical factor $s=1/(1+\delta_{BC})$  is needed only if $B$ and $C$ are identical particles, and always equals to one in present work.

In our calculation, we take the same notations of $\Lambda_b$ and $\Sigma_b$ baryons as those in Ref.~\cite{Chen:2016spr}. The explicit notations of relevant excited bottom baryons together with the predicted masses~\cite{Ebert:2011kk} are listed in Table~\ref{tab1}. For the masses of initial $\Lambda_b(2S)$, $\Lambda_b(1D)$, $\Sigma_b(2S)$, and $\Sigma_b(1P)$ states, we first adopt the masses of $\Lambda_b(6146)^0~[\Sigma_b(6146)^0]$ and $\Lambda_b(6152)^0~[\Sigma_b(6152)^0]$ from LHCb experimental data by assuming that they are possible candidates. If the possible assignment for the $\Lambda_b(6146)^0~[\Sigma_b(6146)^0]$ or $\Lambda_b(6152)^0~[\Sigma_b(6152)^0]$ is finally excluded via decay width, the predicted mass of this state is employed to recalculate its strong decay behaviors. For the final ground states, the masses are taken from the Review of Particle Physics~\cite{Tanabashi:2018oca}. For the harmonic oscillator parameter of $\pi$ meson, the effective value $R= 2.5~\rm{GeV^{-1}}$ is employed~\cite{Godfrey:2015dva}. For the baryon parameters, we use $\alpha_\rho=400~\rm{MeV}$ and
\begin{eqnarray}
\alpha_\lambda=\Bigg(\frac{3m_Q}{2m_q+m_Q} \Bigg)^\frac{1}{4} \alpha_\rho,
\end{eqnarray}
where the $m_Q$ and $m_q$ are the heavy and light quark masses, respectively~\cite{Lu:2018utx,Zhong:2007gp}. The $m_{u/d}=220~\rm{MeV}$ and $m_b=4.977~\rm{MeV}$ are introduced to explicitly consider the mass differences between light and bottom quark~\cite{Godfrey:1985xj,Capstick:1986bm,Godfrey:2015dva}.
The overall parameter $\gamma$, can be determined by the well established $\Sigma_c(2520)^{++} \to \Lambda_c \pi^+$
process. The $\gamma=9.83$ is obtained by reproducing the width $\Gamma [\Sigma_c(2520)^{++} \to \Lambda_c \pi^+]=14.78~\rm{MeV}$~\cite{Tanabashi:2018oca,Lu:2018utx}. With these parameters, the calculated decay widths of the ground states $\Sigma_b$ and $\Sigma_b^*$ to $\Lambda_b \pi$ are shown in Table~\ref{ground}, which agree with the experimental data well.

\begin{table*}[!htbp]
\begin{center}
\caption{ \label{tab1} Notations, quantum numbers, and the predicted masses of the relevant excited singly bottom baryons. For the spatial $2S$ excited states, the symbol $2S$ are added in the brackets. The $n_\rho$ and $L_\rho$ stand for the radial and orbital angular momenta between the two light quarks, respectively, while $n_\lambda$ and $L_\lambda$ denote the radial and angular momenta between the two light quark system and the bottom quark, respectively. $L$ is the total orbital angular momentum, $S_\rho$ is the total spin of the light quarks, $j$ is total angular momentum of $L$ and $S_\rho$, $J$ is the total angular momentum, and $P$ corresponds to the parity of the hadron. The predicted masses of these excited states are taken from the relativistic quark model~\cite{Ebert:2011kk}. The units are in MeV.}
\renewcommand{\arraystretch}{1.5}
\normalsize
\begin{tabular*}{18cm}{@{\extracolsep{\fill}}*{10}{p{1.5cm}<{\centering}}}
\hline\hline
 State                          & $n_\rho$ & $L_\rho$     &  $n_\lambda$      &  $L_\lambda$  &  $L$  & $S_\rho$ & $j$  & $J^P$  & Mass~\cite{Ebert:2011kk}       \\\hline
 $\Lambda_b(2S)$     & 0        & 0            &  1                &  0            &  0    & 0        & 0      & $\frac{1}{2}^+$    & 6089 \\
 $\Lambda_{b2}(\frac{3}{2}^+)$     & 0        & 0            &  0                &  2            &  2    & 0        & 2      & $\frac{3}{2}^+$   & 6190 \\
 $\Lambda_{b2}(\frac{5}{2}^+)$     & 0        & 0            &  0                &  2            &  2    & 0        & 2      & $\frac{5}{2}^+$   & 6196 \\
 $\Sigma_b(2S)$      & 0        & 0            &  1                &  0            &  0    & 1        & 1      & $\frac{1}{2}^+$   & 6213\\
 $\Sigma_b^*(2S)$    & 0        & 0            &  1                &  0            &  0    & 1        & 1      & $\frac{3}{2}^+$  & 6226\\
 $\Sigma_{b0}(\frac{1}{2}^-)$     & 0        & 0            &  0                &  1            &  1    & 1        & 0      & $\frac{1}{2}^-$   & 6101 \\
 $\Sigma_{b1}(\frac{1}{2}^-)$     & 0        & 0            &  0                &  1            &  1    & 1        & 1      & $\frac{1}{2}^-$   & 6095 \\
 $\Sigma_{b1}(\frac{3}{2}^-)$     & 0        & 0            &  0                &  1            &  1    & 1        & 1      & $\frac{3}{2}^-$  & 6096 \\
 $\Sigma_{b2}(\frac{3}{2}^-)$     & 0        & 0            &  0                &  1            &  1    & 1        & 2      & $\frac{3}{2}^-$  & 6087 \\
 $\Sigma_{b2}(\frac{5}{2}^-)$     & 0        & 0            &  0                &  1            &  1    & 1        & 2      & $\frac{5}{2}^-$   & 6084 \\
 \hline\hline
\end{tabular*}
\end{center}
\end{table*}

\begin{table}[!htbp]
\begin{center}
\caption{\label{ground} Strong decays of the ground states $\Sigma_b$ and $\Sigma_b^*$ to $\Lambda_b \pi$ together with experimental data in MeV.}
\renewcommand{\arraystretch}{1.5}
\normalsize
\begin{tabular*}{8.5cm}{@{\extracolsep{\fill}}*{5}{p{1.5cm}<{\centering}}}
\hline\hline
Mode	&	$\Sigma_b^+$       &  $\Sigma_b^-$  &$\Sigma_b^{*+}$       &  $\Sigma_b^{*-}$	\\\hline
$\Lambda_b \pi$	&	5.08	     & 6.11    &9.71   &10.99\\
Experiments 	&	$5.0\pm0.5$ & $5.3\pm0.5$&	$9.4\pm0.5$ & $10.4\pm0.8$ \\
\hline\hline
\end{tabular*}
\end{center}
\end{table}

\section{Strong decay}{\label{decay}}

\subsection{$\Lambda_b(2S)$ state}

In the conventional quark model, only one $\lambda-$mode $\Lambda_b(2S)$ state exists. From Table~\ref{tab1}, the predicted mass of $\Lambda_b(2S)$ state within the relativistic quark model is about 6089 MeV. Considering the uncertainties of different quark models, it is possible to assign the newly observed $\Lambda_b(6146)^0$ or $\Lambda_b(6152)^0$ as the $\Lambda_b(2S)$ state. The strong decay behaviors of the $\Lambda_b(2S)$ assignments are listed in Table~\ref{lamb2s1}. It is shown that the total decay widths of the $\Lambda_b(6146)^0$ and $\Lambda_b(6152)^0$ as $\Lambda_b(2S)$ state are 28.23 MeV and 30.28 MeV, respectively. Compared with the experimental total widths $2.9\pm1.3\pm0.3~\rm{MeV}$ and $2.1\pm0.8\pm0.3~\rm{MeV}$ measured by LHCb Collaboration, both assignments can be excluded.

\begin{table}[!htbp]
\begin{center}
\caption{\label{lamb2s1} Strong decays of the $\Lambda_b(6146)^0$ and $\Lambda_b(6152)^0$ as $\Lambda_b(2S)$ states in MeV.}
\renewcommand{\arraystretch}{1.5}
\normalsize
\begin{tabular*}{8.5cm}{@{\extracolsep{\fill}}*{3}{p{2.5cm}<{\centering}}}
\hline\hline
Decay	&	\multicolumn{2}{c}{$\Lambda_b(2S)$}    \\
Mode	&	$\Lambda_b(6146)^0$       &  $\Lambda_b(6152)^0$ 	\\\hline
$\Sigma_b^+ \pi^-$	&	3.69	     & 3.94  \\
$\Sigma_b^0 \pi^0$	&	3.75	     & 3.99\\
$\Sigma_b^- \pi^+$	&	3.52	     & 3.77\\
$\Sigma_b^{*+} \pi^-$	&	5.79	 & 6.23\\
$\Sigma_b^{*0} \pi^0$	&	5.9	     & 6.34\\
$\Sigma_b^{*-} \pi^+$	&	5.58	 & 6.01\\
Total width	&	28.23                & 30.28 \\
Experiments 	&	$2.9\pm1.3\pm0.3$ & $2.1\pm0.8\pm0.3$ \\
\hline\hline
\end{tabular*}
\end{center}
\end{table}

From the heavy quark flavor symmetry, the mass gap between $\Lambda_b(2S)$ and $\Lambda_b$ should be similar with the charmed baryon case, that is
\begin{eqnarray}
m[\Lambda_b(2S)]- m[\Lambda_b]  \approx  m[\Lambda_c(2S)]- m[\Lambda_c] = 480~\rm{MeV}.
\end{eqnarray}
Hence, the mass of $\Lambda_b(2S)$ is about 6100 MeV, which is consistent with the quark model prediction. The strong decays of $\Lambda_b(2S)$ with a mass of 6089 MeV are presented in Table~\ref{lamb2s2}. It can be seen that $\Lambda_b(2S)$ state is relatively narrow due to the small phase space. The partial decay width ratios of the $\Lambda_b(2S)$ state are predicted to be
\begin{eqnarray}
\Gamma[\Sigma_b \pi]:\Gamma[\Sigma_b^* \pi]  =0.70,
\end{eqnarray}
which is independent with the overall parameter $\gamma$ in the $^3P_0$ model. For the total decay width of $\Lambda_b(2S)$, our result is about three times smaller than that of Ref.~\cite{Chen:2018vuc}. However, our predicted branching ratios agree with the results of Ref~\cite{Chen:2018vuc}, which can provide useful information to search for the $\Lambda_b(2S)$ in future experiments.

\begin{table}
\begin{center}
\caption{\label{lamb2s2} Strong decays of the $\Lambda_b(2S)$ state with a mass of 6089 MeV.}
\renewcommand{\arraystretch}{1.5}
\normalsize
\begin{tabular*}{8.5cm}{@{\extracolsep{\fill}}*{2}{p{3.5cm}<{\centering}}}
\hline\hline
Mode	&	$\Lambda_b(2S)$     	\\\hline
$\Sigma_b^+ \pi^-$	&	1.75	     \\
$\Sigma_b^0 \pi^0$	&	1.80	    \\
$\Sigma_b^- \pi^+$	&	1.64	     \\
$\Sigma_b^{*+} \pi^-$	&	2.47	 \\
$\Sigma_b^{*0} \pi^0$	&	2.56	 \\
$\Sigma_b^{*-} \pi^+$	&	2.34	 \\
Total width	&	12.56                 \\
\hline\hline
\end{tabular*}
\end{center}
\end{table}

\subsection{$\Lambda_b(1D)$ states}

There are two $\Lambda_b(1D)$ states in the constituent quark model, which form a doublet. From Table~\ref{tab1}, the predicted masses of the two $\Lambda_b(1D)$ states are 6190 MeV and 6196 MeV, respectively. The predicted masses and small mass splitting suggest that the $\Lambda_b(6146)^0$ and $\Lambda_b(6152)^0$ are good candidates of this $\Lambda_b(1D)$ doublet. Although the predicted mass of the $J^P=3/2^+$ state is slightly smaller than that of the $J^P=5/2^+$ state with a normal mass order, the inversed mass order case is also possible physically. Hence, we consider both cases to fully examine the possible assignments. Their strong decays of the $\Lambda_b(1D)$ assignments are shown in Table~\ref{lamb1d}. The predicted total decay widths of these assignments are several MeV, which indicate the two $\Lambda_b(1D)$ states are both narrow states. The strong decays of these $\Lambda_b(1D)$ states are also investigated within the chiral quark model~\cite{Yao:2018jmc}, which are consistent with our present results. Compared with the experimental data, our theoretical results seem to be larger. However, considering the uncertainties of the $^3P_0$ model and experiments, these assignments are acceptable.

\begin{table}
\begin{center}
\caption{\label{lamb1d} Strong decays of the $\Lambda_b(6146)^0$ and $\Lambda_b(6152)^0$ as $\Lambda_b(1D)$ states in MeV.}
\renewcommand{\arraystretch}{1.5}
\small
\begin{tabular*}{8.8cm}{@{\extracolsep{\fill}}*{5}{p{1.6cm}<{\centering}}}
\hline\hline
Decay	&	\multicolumn{2}{c}{$\Lambda_{b2}(\frac{3}{2}^+)$} & \multicolumn{2}{c}{$\Lambda_{b2}(\frac{5}{2}^+)$}  \\
Mode	&	$\Lambda_b(6146)^0$       &  $\Lambda_b(6152)^0$  &$\Lambda_b(6146)^0$       &  $\Lambda_b(6152)^0$	\\\hline
$\Sigma_b^+ \pi^-$	&	1.79	     & 1.91    &0.01   &0.01\\
$\Sigma_b^0 \pi^0$	&	1.82	     & 1.94    &0.01   &0.01\\
$\Sigma_b^- \pi^+$	&	1.70	     & 1.82    &0.01   &0.01\\
$\Sigma_b^{*+} \pi^-$	&	0.29	 & 0.31    &1.68   &1.81\\
$\Sigma_b^{*0} \pi^0$	&	0.30	 & 0.32    &1.72   &1.85\\
$\Sigma_b^{*-} \pi^+$	&	0.28	 & 0.30    & 1.62  &1.75\\
Total width	&	6.17                & 6.60     & 5.05  & 5.44\\
Experiments 	&	$2.9\pm1.3\pm0.3$ & $2.1\pm0.8\pm0.3$	&$2.9\pm1.3\pm0.3$ &$2.1\pm0.8\pm0.3$ \\
\hline\hline
\end{tabular*}
\end{center}
\end{table}

\begin{figure*}[htb]
\includegraphics[scale=0.60]{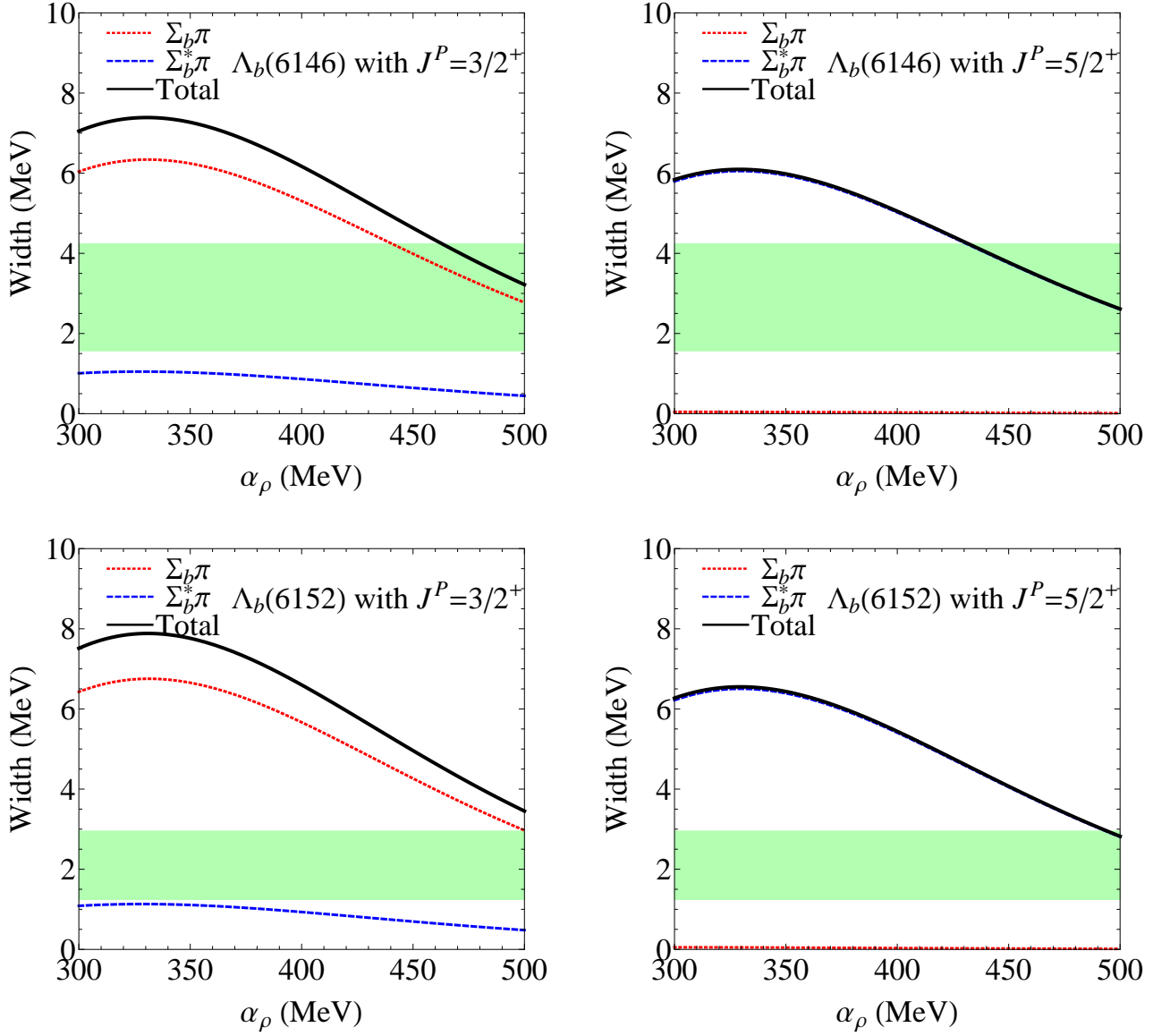}
\vspace{0.0cm} \caption{The dependence on the harmonic oscillator parameter $\alpha_\rho$ of the initial states. The green bands stand for the measured total decay widths with errors.}
\label{lambdab1d}
\end{figure*}

Two factors in the quark pair creation model can affect the calculated total decay widths: one is the harmonic oscillator parameter $\alpha_\rho$, and the other is the overall strength $\gamma$. The dependence on the harmonic oscillator parameter $\alpha_\rho$ of the initial states is shown in Fig.~\ref{lambdab1d}.  When the $\alpha_\rho$ increases, the total decay widths decrease in a wide parameter range and the total decay widths remain to be several MeV. Within the reasonable range of the parameter $\alpha_\rho$, our results suggest that the $\Lambda_b(6146)^0$ and $\Lambda_b(6152)^0$ can be clarified into the $\Lambda_b(1D)$ doublet.

The uncertainties arising from quark pair creation strength $\gamma$ can be eliminated if we only concentrate on the branching ratios of $\Sigma_b \pi$ and $\Sigma_b^* \pi$ channels. Since $\Lambda_{b2}(\frac{3}{2}^+)$ and $\Lambda_{b2}(\frac{5}{2}^+)$ belong to the same $j=2$ doublet, the heavy quark spin symmetry suggest that they may have the similar properties, such as masses and total decay widths. The ratios between $\Sigma_b \pi$ and $\Sigma_b^* \pi$ channels decay modes may help us to distinguish these two states. For the $J^P=3/2^+$ state, the partial decay width ratios are predicted to be
\begin{eqnarray}
\Gamma[\Lambda_b(6146)^0 \to \Sigma_b \pi]:\Gamma[\Lambda_b(6146)^0 \to \Sigma_b^* \pi]  = 6.14,
\end{eqnarray}
\begin{eqnarray}
\Gamma[\Lambda_b(6152)^0 \to \Sigma_b \pi]:\Gamma[\Lambda_b(6152)^0 \to \Sigma_b^* \pi]  = 6.07.
\end{eqnarray}
For the $J^P=5/2^+$ state, the partial decay width ratios are predicted to be
\begin{eqnarray}
\Gamma[\Lambda_b(6146)^0 \to \Sigma_b \pi]:\Gamma[\Lambda_b(6146)^0 \to \Sigma_b^* \pi]  = 6\times10^{-3},
\end{eqnarray}
\begin{eqnarray}
\Gamma[\Lambda_b(6152)^0 \to \Sigma_b \pi]:\Gamma[\Lambda_b(6152)^0 \to \Sigma_b^* \pi]  = 6\times10^{-3}.
\end{eqnarray}
It can be seen that the branching ratio of $\Sigma_b \pi$ channel is rather small for the $J^P=5/2^+$ state, which may be hardly seen in experiments.  Experimentally, $\Lambda_b(6146)^0 \to \Sigma_b^* \pi$, $\Lambda_b(6152)^0 \to \Sigma_b \pi$, and $\Lambda_b(6152)^0 \to \Sigma_b^* \pi$ processes have been clearly visible, while the $\Lambda_b(6146)^0 \to \Sigma_b \pi$ mode is no significant evidence~\cite{lhcb}. This suggests that $\Lambda_b(6146)^0$ and $\Lambda_b(6152)^0$ may favor the $J^P=5/2^+$ and $J^P=3/2^+$ states, respectively. More information on masses, partial decay widths, and spins are needed to test our present assignments.

\subsection{$\Sigma_b(2S)$ states}

From Table~\ref{tab1}, the predicted masses of $\Sigma_b(2S)$ and $\Sigma_b^*(2S)$ are 6213 MeV and 6226 MeV, respectively. By assuming the two resonances are $\Sigma_b$ states, the strong decays are calculated and listed in Table~\ref{sigmab2s1}. Compared with experimental data, these assignments can be excluded.
\begin{table}
\begin{center}
\caption{\label{sigmab2s1} Strong decays of the $\Sigma_b(6146)$ and $\Sigma_b(6152)$ as $\Sigma_b(2S)$ and $\Sigma_b^*(2S)$ states in MeV.}
\renewcommand{\arraystretch}{1.5}
\small
\begin{tabular*}{8.8cm}{@{\extracolsep{\fill}}*{5}{p{1.6cm}<{\centering}}}
\hline\hline
Decay	&	\multicolumn{2}{c}{$\Sigma_{b}(2S)$} & \multicolumn{2}{c}{$\Sigma_{b}^*(2S)$}  \\
Mode	&	$\Sigma_b(6146)$       &  $\Sigma_b(6152)$  &$\Sigma_b(6146)$       &  $\Sigma_b(6152)$	\\\hline
$\Lambda_b \pi$	&	13.55	     & 13.85    &13.55   &13.85\\
$\Sigma_b \pi$	&	9.92	     & 10.57    &2.48   &2.64\\
$\Sigma_b^* \pi$	&	3.90	 & 4.19    &9.75   &10.48\\
Total width	&	27.37                & 28.62     & 25.78  & 26.98\\
Experiments 	&	$2.9\pm1.3\pm0.3$ & $2.1\pm0.8\pm0.3$	&$2.9\pm1.3\pm0.3$ &$2.1\pm0.8\pm0.3$ \\
\hline\hline
\end{tabular*}
\end{center}
\end{table}

With the predicted masses of $\Sigma_b(2S)$ and $\Sigma_b^*(2S)$, their strong decays are presented in Table~\ref{sigmab2s2}. Our results show that the masses and total decay widths of $\Sigma_b(2S)$ and $\Sigma_b^*(2S)$ states are similar. For the $\Sigma_b(2S)$ state, the branching ratios are
\begin{eqnarray}
Br(\Lambda_b \pi, \Sigma_b \pi, \Sigma_b^* \pi)  = 39\%, 43\%, 18\%.
\end{eqnarray}
For the $\Sigma_b^*(2S)$ state, the branching ratios are
\begin{eqnarray}
Br(\Lambda_b \pi, \Sigma_b \pi, \Sigma_b^* \pi)  = 39\%, 12\%, 49\%.
\end{eqnarray}
These branching ratios are independent with the overall constant $\gamma$ and can be adopted to distinguish these two states. Also, our predicted total widths of $\Sigma_b(2S)$ and $\Sigma_b^*(2S)$ states are smaller than that of Ref.~\cite{Chen:2018vuc}, while the branching ratios are similar.

\begin{table}
\begin{center}
\caption{\label{sigmab2s2} Strong decays of the $\Sigma_b(2S)$ and $\Sigma_b^*(2S)$ states with predicted masses.}
\renewcommand{\arraystretch}{1.5}
\normalsize
\begin{tabular*}{8.5cm}{@{\extracolsep{\fill}}*{3}{p{2.5cm}<{\centering}}}
\hline\hline
Mode	&	$\Sigma_b(2S)$       &  $\Sigma_b^*(2S)$ 	\\\hline
$\Lambda_b \pi$	&	16.50	     & 16.94   \\
$\Sigma_b \pi$	&	18.29	     & 5.03    \\
$\Sigma_b^* \pi$&   7.74	 & 21.53    \\
Total width	&	42.53         & 43.50   \\
\hline\hline
\end{tabular*}
\end{center}
\end{table}

\subsection{$\Sigma_b(1P)$ states}

There are five $\lambda-$mode $\Sigma_b(1P)$ states, denoted as $\Sigma_{b0}(\frac{1}{2}^-)$, $\Sigma_{b1}(\frac{1}{2}^-)$, $\Sigma_{b1}(\frac{3}{2}^-)$, $\Sigma_{b2}(\frac{3}{2}^-)$, and $\Sigma_{b2}(\frac{5}{2}^-)$, respectively. The strong decays of $\Sigma_b(6146)$ and $\Sigma_b(6152)$ as these states are listed in Table~\ref{sigmab1p1}. It can be seen that the $\Sigma_b(6146)$ and $\Sigma_b(6152)$ as pure $\Sigma_b(1P)$ assignments can be excluded.

\begin{table*}
\begin{center}
\caption{ \label{sigmab1p1} Decay widths of the $\Sigma_b(6146)$ and $\Sigma_b(6152)$ as $\Sigma_b(1P)$ states in MeV.}
\renewcommand{\arraystretch}{1.5}
\footnotesize
\begin{tabular*}{18cm}{@{\extracolsep{\fill}}*{11}{p{1.5cm}<{\centering}}}
\hline\hline
 Decay     & \multicolumn{2}{c}{$\Sigma_{b0}(\frac{1}{2}^-)$}   & \multicolumn{2}{c}{$\Sigma_{b1}(\frac{1}{2}^-)$} & \multicolumn{2}{c}{$\Sigma_{b1}(\frac{3}{2}^-)$}  &  \multicolumn{2}{c}{$\Sigma_{b2}(\frac{3}{2}^-)$} & \multicolumn{2}{c}{$\Sigma_{b2}(\frac{5}{2}^-)$} \\
 Mode     & $\Sigma_b(6146)$  & $\Sigma_b(6152)$ & $\Sigma_b(6146)$ & $\Sigma_b(6152)$ & $\Sigma_b(6146)$ & $\Sigma_b(6152)$ &  $\Sigma_b(6146)$ & $\Sigma_b(6152)$& $\Sigma_b(6146)$ & $\Sigma_b(6152)$\\
 $\Lambda_b \pi$	&	247.60	 & 245.01  & $-$    & $-$     &$-$    & $-$   &  19.21 & 20.34   & 19.21  &20.34\\
$\Sigma_b \pi$	    &	$-$	     & $-$     & 294.79 & 300.33  &1.23   &1.37   &  2.20  & 2.47   & 0.98    &1.10\\
$\Sigma_b^* \pi$    &   $-$	     & $-$     & 1.62   & 1.83    &274.59 &281.07 &  1.45  & 1.65   & 2.26    &2.56\\
 Total              & 247.60     & 245.01  & 296.41 & 302.16  &275.81 &282.44 &  22.87 & 24.46   & 22.45  &24.00 \\
 Experiments 	    &	$2.9\pm1.3\pm0.3$ & $2.1\pm0.8\pm0.3$	&$2.9\pm1.3\pm0.3$ &$2.1\pm0.8\pm0.3$ &$2.9\pm1.3\pm0.3$ &$2.1\pm0.8\pm0.3$ &$2.9\pm1.3\pm0.3$ &$2.1\pm0.8\pm0.3$ &$2.9\pm1.3\pm0.3$ &$2.1\pm0.8\pm0.3$\\
\hline\hline
\end{tabular*}
\end{center}
\end{table*}

In the literature, the $\Sigma_b(6097)$ observed by LHCb Collaboration~\cite{Aaij:2018tnn} are clarified into the $\Sigma_b(1P)$ states~\cite{Chen:2018vuc,Wang:2018fjm,Yang:2018lzg,Aliev:2018vye,Cui:2019dzj,Jia:2019bkr}. Here, we present the our calculations of $\Sigma_b(6097)$ as $\Sigma_b(1P)$ states in the Table~\ref{sigmab1p2}. Our results do not favor the $\Sigma_b(6097)$ as the pure $\Sigma_b(1P)$ states. However, the $\Sigma_b(6097)$ may correspond to the mixing state of $\Sigma_{b1}(\frac{3}{2}^-)$ and $\Sigma_{b2}(\frac{3}{2}^-)$ states.

\begin{table}
\begin{center}
\caption{ \label{sigmab1p2} Decay widths of the $\Sigma_b(6097)$  as $\Sigma_b(1P)$ states in MeV.}
\renewcommand{\arraystretch}{1.5}
\normalsize
\begin{tabular}{lcccccccccc}
\hline\hline
 Mode     & $\Sigma_{b0}(\frac{1}{2}^-)$   & $\Sigma_{b1}(\frac{1}{2}^-)$ & $\Sigma_{b1}(\frac{3}{2}^-)$  &  $\Sigma_{b2}(\frac{3}{2}^-)$ & $\Sigma_{b2}(\frac{5}{2}^-)$    \\
 $\Lambda_b \pi$	&	261.68	   & $-$    & $-$       & 11.54  &11.54   \\
$\Sigma_b \pi$	    &	$-$	       & 240.92 & 0.43      & 0.77   & 0.34   \\
$\Sigma_b^* \pi$    &   $-$	       & 0.49   & 214.45    & 0.45   & 0.69   \\
 Total              &   261.68     & 241.41 & 214.88    & 12.75   & 12.57   \\
 Experiments 	    &	\multicolumn{5}{c}{$31.0\pm5.5\pm0.7$} \\
\hline\hline
\end{tabular}
\end{center}
\end{table}

\begin{figure}[htb]
\includegraphics[scale=0.27]{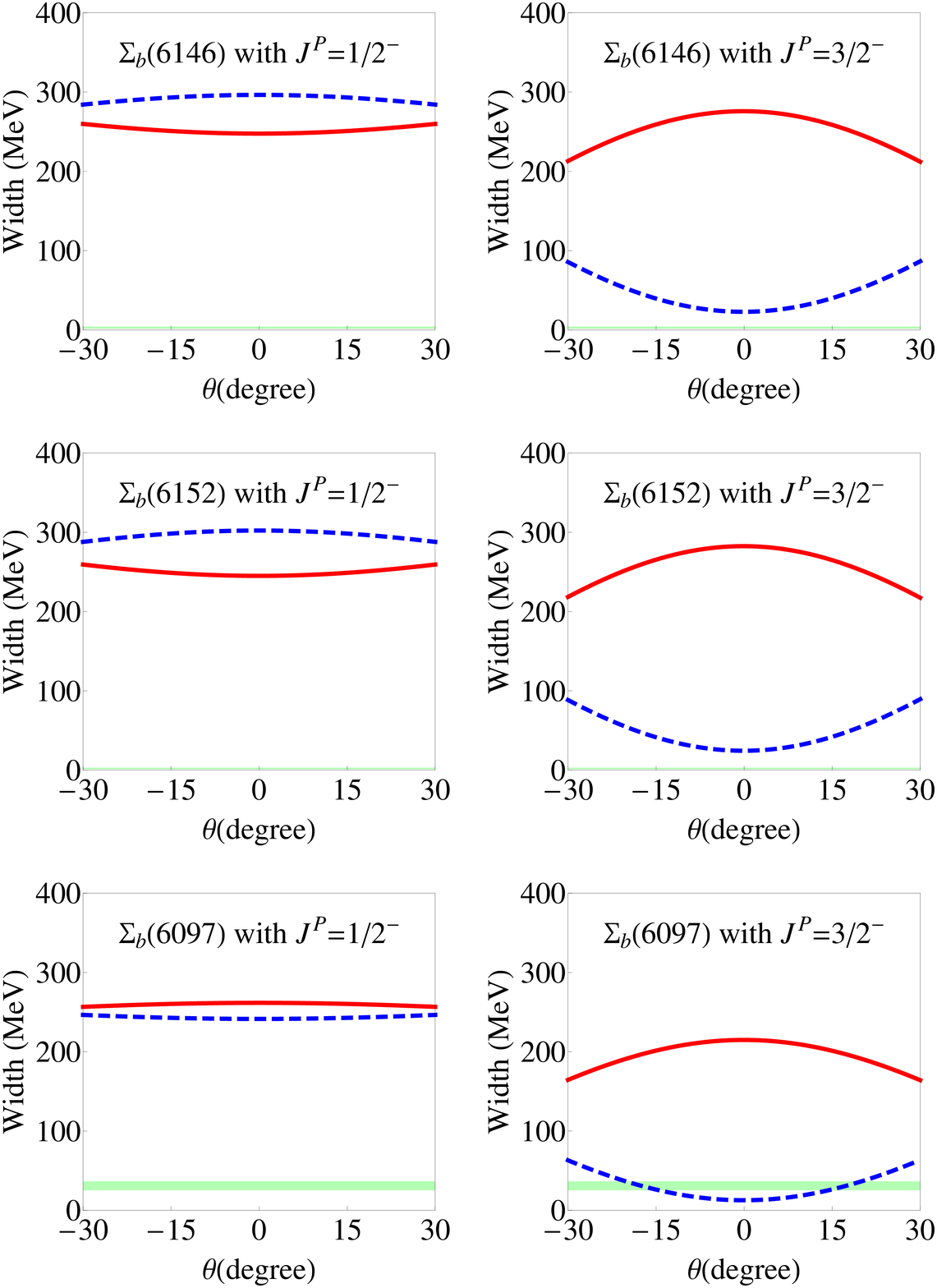}
\vspace{0.0cm} \caption{The total decay widths of various assignments as functions of the mixing angle $\theta$ in the range $-30^\circ \sim 30^\circ$. The red solid lines are the $| P~{3/2^-}\rangle_1$ states, and the blue dashed curves correspond to the $| P~{3/2^-}\rangle_2$ states. The green bands stand for the measured total decay widths with errors.}
\label{sigmab1p}
\end{figure}

For completeness, we investigate the possibilities of $\Sigma_b(6146)$, $\Sigma_b(6152)$ and $\Sigma_b(6097)$ as the $P-$wave mixing states. The physical states can be the mixing of the quark model states with the same $J^P$, that is
\begin{equation}
\left(\begin{array}{c}| P~{1/2^-}\rangle_1\cr |  P~{1/2^-}\rangle_2
\end{array}\right)=\left(\begin{array}{cc} \cos\theta & \sin\theta \cr -\sin\theta &\cos\theta
\end{array}\right)
\left(\begin{array}{c} |1/2^-,j=0
\rangle \cr |1/2^-,j=1\rangle
\end{array}\right),
\end{equation}
\begin{equation}
\left(\begin{array}{c}| P~{3/2^-}\rangle_1\cr |  P~{3/2^-}\rangle_2
\end{array}\right)=\left(\begin{array}{cc} \cos\theta & \sin\theta \cr -\sin\theta &\cos\theta
\end{array}\right)
\left(\begin{array}{c} |3/2^-,j=1
\rangle \cr |3/2^-,j=2\rangle
\end{array}\right).
\end{equation}
In the heavy quark limit, the mixing angle should equal to zero. Given the finite mass of bottom quark, the mixing angle may have small divergence between the physical states and the $j-j$ coupling scheme to approximately preserve the heavy quark symmetry. The total decay widths of various assignments versus the mixing angle $\theta$ in the range $-30^\circ \sim 30^\circ$ are plotted in Figure~\ref{sigmab1p}. It is shown that the $\Sigma_b(6146)$ and $\Sigma_b(6152)$ as the $\Sigma_b(1P)$ states can be fully excluded. With the mixing angle $\theta \cong 17^\circ$, the $\Sigma_b(6097)$ can be interpreted as the $J^P=3/2^-$ mixing state, where the $j=2$ component dominate. This assignment also agrees with other theoretical works~\cite{Chen:2018vuc,Wang:2018fjm,Yang:2018lzg,Aliev:2018vye,Cui:2019dzj,Jia:2019bkr}. The the branching ratios are
\begin{eqnarray}
Br(\Lambda_b \pi, \Sigma_b \pi, \Sigma_b^* \pi)  = 35.0\%, 3.5\%, 61.5\%,
\end{eqnarray}
which are independent with the overall parameter $\gamma$ and can be tested by the future experiments.

\section{Summary}{\label{Summary}}

 In this work, we study the strong decay behaviors of the newly observed two resonances $\Lambda_b(6146)^0~[\Sigma_b(6146)^0]$ and $\Lambda_b(6152)^0~[\Sigma_b(6152)^0]$ by LHCb Collaboration. Given their masses and decay modes, the two resonances can be tentatively assigned as the $\lambda-$ mode $\Lambda_b(2S)$, $\Lambda_b(1D)$, $\Sigma_b(2S)$, and $\Sigma_b(1P)$ states. Their strong decay behaviors are calculated with the quark pair creation model. Compared with the experimental data measured by LHCb Collaboration, the $\Lambda_b(6146)^0$ and $\Lambda_b(6152)^0$ can be clarified into the $\Lambda_b(1D)$ doublet reasonably, while other canonical assignments are disfavored.

 For the $J^P=3/2^+$ state, the partial decay width ratios of $\Sigma_b \pi$ and $\Sigma_b^* \pi$ channels are about 6, while the ratios are about $6\times 10^{-3}$ for the $J^P=5/2^+$ state. This indicates that the $\Sigma_b \pi$ channel may be hardly observed for the $J^P=5/2^+$ state experimentally. Together with experimental decay modes, the $\Lambda_b(6146)^0$ and $\Lambda_b(6152)^0$ may favor the $J^P=5/2^+$ and $J^P=3/2^+$ states, respectively. More experimental data on masses, partial decay widths, and spins are needed to test our present assignments. Other investigations of the low-lying bottom states $\Lambda_b(2S)$, $\Sigma_b(2S)$, and $\Sigma_b(1P)$ can also provide helpful information for future experimental searches.

\bigskip
\noindent
\begin{center}
{\bf ACKNOWLEDGEMENTS}\\

\end{center}
This project is supported by the National Natural Science Foundation of China under Grants No.~11705056, No.~11775078, and No.~U1832173.

\end{document}